# The Jodcast

N. J. Rattenbury

Jodrell Bank Centre for Astrophysics, The University of Manchester

**Summary**. The Jodcast (www.jodcast.net) is a twice-monthly astronomy podcast from The University of Manchester's Jodrell Bank Observatory. In this paper I give the motivation and history of The Jodcast, as well as a description of The Jodcast's content, operations, personnel, performance and aspirations.

## 1 Introduction

Every day during the school vacations, visitors to Jodrell Bank Observatory gather around a volunteer JBO astronomer to listen to a short history of the Observatory, followed by an opportunity to ask their questions on space and astronomy. This is the "Ask-An-Astronomer" session, and usually draws crowds of 20-30 people.

Following one such session, a debate ensued in the corridor of the JBO control building on whether such a modest audience size justified the time spent by a research-grade astronomer to perform this public outreach service. The question became one of audience size: by what means can such science outreach be done effectively and efficiently to as large an audience as possible?

Creating a podcast appeared the logical answer to this question, and had been informally discussed previously amongst several people at JBO. A podcast is a piece of audio/video or other content "broadcast" over the internet. Users from anywhere in the world can subscribe to the content, downloading new content automatically as it becomes available. Podcasting is rapidly becoming a routine form of media delivery for all broadcasters; mainstream or otherwise.

In the latter half of 2005, plans were drawn up to produce an astronomy podcast from Jodrell Bank Observatory: The Jodcast.

## 2. The Jodcast: Description

### 2.1 Aims and Objectives

The aims of The Jodcast are:
   (i) To inspire and inform people about astronomy and related sciences,
   (ii) To excite young people with the latest results of astronomy research,
   (iii) To break down the stereotypes of science and scientists as being incomprehensible and unapproachable,
   (iv) To become a regular source of astronomy information,
   (v) To motivate students to study science.

### 2.2 Content

Over two episodes each month, The Jodcast presents the latest news from the world of astronomy, including exciting research results, events (both terrestrial and astronomical) and issues relevant to those interested in astronomy. A regular segment on what can be seen in the night sky that month details what objects can be seen with the naked eye, binoculars or a small telescope. Questions from listeners about space and astronomy are answered in the Ask-An-Astronomer segment. In addition to these regular segments of The Jodcast, special interviews with leading researchers are presented.

The Jodcast News is compiled largely from existing media, including print journals such as Nature and Science. Press releases distributed via the web, and associated internet resources are also used to compile the news. Current affairs, not necessarily based on astronomy research, but which affect the astronomy community in general, are also reported.

Content for the Night Sky segment is taken from the existing Night Sky webpages on the JBO website. The emphasis is on what can be seen without instrumentation and features special objects or events each month, such as visible planets or meteor showers. Recently, we have started to include some comments on the Southern hemisphere night sky, following several requests from listeners.

Essentially the genesis of The Jodcast, the Ask-An-Astronomer segment features questions from listeners put to a JBO astronomer. The questions are collected via The Jodcast feedback webpage, or from the occasional letter. Many questions are on topics which involve complicated physics and their answers require careful preparation and research.

The original intention of the Ask-An-Astronomer segment of The Jodcast was to answer all listener questions via return email, and only broadcast the most interesting questions and their answers. Typically, the quality of questions and the level of interest each corresponding answer would hold for a general audience is such that we have tended to broadcast the large majority of received questions and their answers. We have however, started to refer listeners who have asked a question which has been answered before to the relevant show in the archive.

The Ask-An-Astronomer segment engages the audience with the opportunity to get involved with the science presented in all parts of The Jodcast. Listeners frequently ask questions on items mentioned in the news, the observation of the night sky – particularly specific objects, or on topics covered by the special interviews. Current affairs from the wider science world are also mentioned, such as the recent interest in the Large Hadron Collider experiment.

In almost every episode of The Jodcast, we feature an extended interview with an astronomer, usually on their current topic of research. Past interviews covered topics ranging across all aspects of astronomy, astrophysics and space science; covering distance scales from the microscopic to the cosmic. Research driven by data at all observation wavelengths has been described, from gravitational waves through radio and visible light, to x-rays and gamma rays. Pulsars, black holes, galaxies, extra-solar planets, alien life, exploding stars – The Jodcast has covered and continues to follow, the most exciting astronomy research and discoveries.

The interviewees are usually academic visitors to Jodrell Bank Observatory, and/or The Jodrell Bank Centre for Astrophysics. The interviewees are usually either visiting Manchester for the purpose of giving one of the regular academic research seminars, or to collaborate with a research colleague. In some cases, our interviewees are in Manchester for a specific event, such as the recent 50$^{th}$ anniversary celebrations at Jodrell Bank Observatory, or to attend a research conference.

Interviews have been conducted with staff from or visitors to, other University of Manchester Schools or departments. The Jodcast has rarely travelled for the explicit purpose of obtaining an audio interview. This is largely due to the relatively large amount of time required to travel to obtain interviews from remote sources. On a few occasions, when a particularly high profile result has been an-nounced, the interview was conduced over the phone, or via Skype.

The key to the success of the interview segments are The Jodcast policies regarding the content and presentation of the science in each interview. We allow the scientists to tell their own story of their research, with as little interruption from the interviewer as possible. We also allow the story to be told at the rate and at a level of detail determined by the interviewee, with as few comments or questions from the interviewer as possible. There is no script, or list of questions to be answered by the interviewee. Questions, when they are asked, generally seek to clarify a piece of jargon, or to expand further a description of a physical process. This policy of allowing the scientist to tell their own story results in a dialogue that is more akin to a conversation, than an interview. Without set minimum, or maximum time-limits on the interview, the resulting segment frequently runs to over 30 minutes, depending on the ease with which the interviewee has with speaking.

The Jodcast has reported from a number of special events. The Jodcast produced a special show from the UK National Astronomy Meeting, in Preston 2007, and in Belfast 2008. These meetings provided the ideal opportunity to interview a large number of astronomers about their current research. In each case, The Jodcast was afforded press status, allowing access to press facilities.

The Jodcast produced a special show covering the International Astronomical Union General Assembly in Prague in 2006. Again, many interviews were conducted with a wide range of astronomers. Of special interest was the debate over the reclassification of Pluto as a dwarf planet, a topic which aroused intense public interest.

The required level of background knowledge and comprehension for all segments of The Jodcast is set to be at that of a senior high-school physics student, or a first-year undergraduate.

**2.3 Anatomy of a show**

The Jodcast was originally produced as a single show, once a month. Following audience requests, from May 2007, The Jodcast began producing two shows per month. Initially, the intention was to split the content of the single show over two episodes, and not to produce more content than that for a single show per month. However, it became clear that we were able to produce enough content to produce two shows per month, each of a length similar to that of the original, once-monthly show.

The first show of each month starts with a short, light-hearted skit – usually a pastiche or parody of a well-known film or TV series. As an introduction to the episode, this sets an informal tone. In the mid-month episode, a short mathematical question or riddle is posed for the listener.

Each show generally has at least two presenters, who introduce the upcoming segments for each show. Commentary and presenter dialogue is kept to a minimum between segments. Frequently we record this linking dialogue between remote presenters via Skype.

In the first show of the month, the News is the first segment to be presented, followed by a discussion and response to listener feedback. This is followed by either the main interview or the Night Sky segment. Final presenter comments are made, usually with a request for more listener feedback, before concluding with another skit – a continuation of the introduction piece, or the answer to the mathematical question/riddle, depending on the show – first of the month, or the mid-month, "Extra" show.

The Ask-An-Astronomer segment appears in the mid-month "Extra" show, along with another interview and more listener feedback.

**2.4 The Ethos of The Jodcast**

The Jodcast is a production which attempts to inspire and educate a general audience without setting specific learning goals. We attempt to set an informal tone, to allow the science to be described by science experts as a conversation, rather than as a series of sound-bites, unnecessarily edited to maximise stimulation. From the first episode, we determined never to attempt to make the subject matter more appealing to a wider audience by "dumbing-down" the science. As a result of setting our typical level of explanation to the understanding of a high-school physics student, our audience tends to be those who have some kind of science training, or background. We consider the effect of this policy in the discussion below, see Section 8.

We note that The Jodcast frequently receives questions from people who clearly have a wide range of general knowledge and understanding of physics and astronomy. In these cases, the corresponding explanations presented in the Ask-An-Astronomer segment similarly vary in depth, and the required level of current understanding.

**2.5 Operations**

For each episode, an Editor is assigned whose main duty is to collate the audio segments from the recordists responsible for each segment, and to edit the full show. The full list of personnel who regularly perform the same tasks for The Jodcast audio podcast are listed below in Table 1.

Table 1. Personnel and regular duties for The Jodcast audio podcast

| Name | Regular Duties | Situation/Location |
|---|---|---|
| Nicholas Rattenbury | Editor, Interviewer, Presenter, Recordist for Night Sky, AaA | PDRA, JBCA |
| Stuart Lowe | Editor, Presenter, Webmaster, RSS maintenance, Graphic artist | PDRA, JBCA |
| David Ault | Presenter, Intro/Outro skit production and voice talent | Repertory theatre, London/India |
| Megan Argo | Compiler & Presenter: News | PDRA, Curtin University of Technology |
| Roy Smits | Editor, Interviewer | PDRA, JBCA |
| Tim O'Brien | The astronomer in Ask-an-Astronomer | Senior Lecturer, JBCA |
| Ian Morison | Compiler & Presenter: Night Sky | Gresham Professor of Astronomy, JBCA. |

The editing of a full show using edited segments typically takes 3-4 hours. The full show is then encoded at three different bit rates, high, standard and low, each with a different trade-off between sound quality and download size. The individual segments – Night Sky, News, Ask-An-Astronomer and the Interview - are also made available as separate downloads, again in response to listener suggestions. RSS feeds are generated for all the individual segments, and for the show at the various bit rates, using Perl scripts.

Descriptive text for the show and each segment are written for the episode webpage, a list compiled of appropriate links to external web resources, and the show credits. Once the episode webpage is ready, the page is made public.

## 3. Technology

The first episodes of The Jodcast were made using equipment ready-to-hand, with no initial investment into equipment, software or any other expenses. These initial shows were recorded directly onto a laptop using freely-available software, and a small lapel microphone which offered marginally superior audio reproduction over the in-built laptop microphone. The funding awarded to The Jodcast (see Section 4) was spent mainly on purchasing high-quality audio equipment.

### 3.1 Audio Equipment

The Jodcast purchased two professional-grade solid state audio recorders (Marantz PMD 660), each with two XLR input channels offering stereo track recording. The audio is recorded onto a removable 2Gb flash RAM module, which allows the storage of around 6 hours to 3 days' worth of audio, depending on whether the audio is stereo or mono, and whether the audio is in native (i.e. uncompressed) or mp3 format. Having two recorders has proven very useful when The Jodcast has covered special events such as the National Astronomy Meetings, allowing two interviewers to collect material at the same time.

Two dynamic microphones were also purchased (Beyerdynamic 58). Much of the improvement in audio quality can be attributed to the use of these high-quality microphones, which incorporate handling noise reduction technology. Using a recording system with XLR cabling also keeps interference noise to a minimum.

**3.2 Software**

The Jodcast is created using the free audio editing software Audacity, which offers an excellent range of features; more than sufficient for producing a podcast. One addition has been Chris Capel's Dynamic Compressor plugin. This allows the perceived loudness of the audio output to be increased, allowing the podcast to be heard more easily in noisy environments, such as when travelling.

**3.3 Website**

A key element of the success of The Jodcast has been the creation and maintenance of a top-quality website: [www.jodcast.net](www.jodcast.net). The emphasis of the front page of The Jodcast website is on simplicity and clarity, with users presented with the cover art of the current and previous two episodes, linking to each episode's own webpage. Links to the archive webpage, or to the contact and feedback pages for example, appear as tabs at the top of the front page. This clean and uncluttered style does not overwhelm new visitors to the site with an excess of information.

The Jodcast maintains a database of keywords for all episodes. Users can search the archive for Jodcast content by entering keywords into a search box, or by clicking onto an element of an up-to-date tag cloud: a graphical representation of the most-frequently used keywords.

As mentioned previously, The Jodcast seeks to actively interact with its audience, eliciting feedback in every episode. Listeners can provide feedback via the corresponding webpage, as well as send in their questions for the Ask-An-Astronomer segment. Feedback is forwarded to all key members of The Jodcast.

## 4. Funding

The Jodcast won funding for 2007 in the form of a Particle Physics and Astronomy Research Council (PPARC) Small Award for the Public Understanding of Science (£2212), and an Institute of Physics Public Engagement Grant (£1000). This funding was used to purchase the equipment detailed above, to pay for advertising, travel and running costs for The Jodcast for one year.

The Jodcast was awarded a further Small Award (£6000) from the Science and Technology Facilities Council (STFC nee PPARC) to continue to produce The Jodcast, and to extend into video podcasting, see Section 9, below.

## 5. Audience

The Jodcast currently has an audience size of over 2800. This figure is the number of times the full show is downloaded off the server. Figure 1 shows the number of downloads of The Jodcast per day following release.

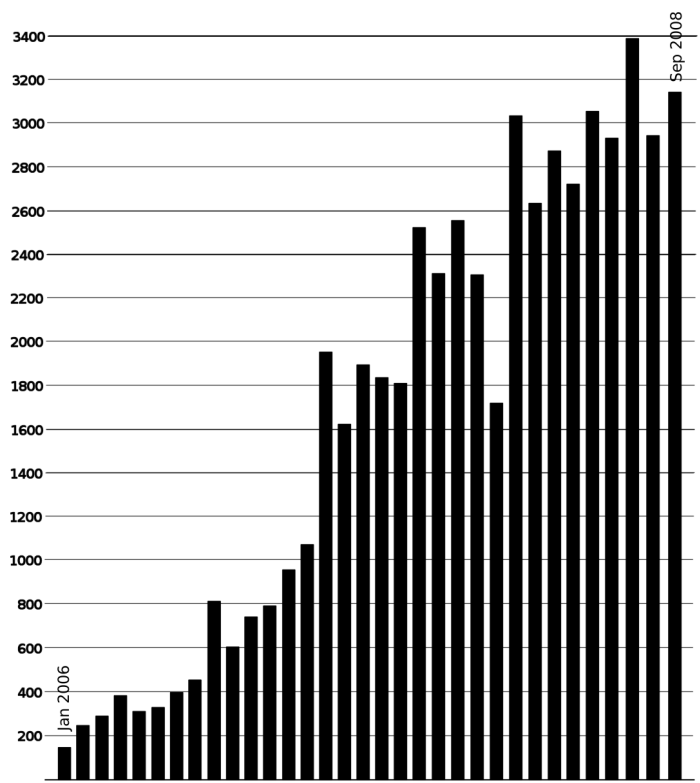

**Fig 1.** Monthly audience size for The Jodcast since the first episode in January 2006, to September 2008.

In 2007, a listener survey was conducted on The Jodcast audience, and the results showed that the survey respondents

(I) are predominately male,
(II) are aged 25 – 70 years old,
(III) live in the UK.

The large majority of the survey respondents found The Jodcast to be of the right length, at the right level and offers interesting information. No survey respondents had an age younger than 25. We might speculate that the format of The Jodcast lends itself more readily to an older audience, however there may be a bias inherent in the age of people more likely to respond to such listener surveys.

Listener feedback to date has been almost completely positive, and all suggestions constructive.

The Jodcast fosters a dialogue with its audience by eliciting and reacting to listener feedback. Suggestions for improving the show, as well as general comments on past items are submitted by listeners via the feedback page of The Jodcast website. As part of each audio podcast episode, listener feedback is read out and comments made on listener suggestions. Where possible, improvements have been made along the lines suggested by listeners. Some suggestions, such as making The Jodcast a weekly podcast, have been rejected for practical reasons.

A group has been created for The Jodcast on the social networking website Facebook. While this allows some opportunity for listeners to interact online, The Jodcast will shortly be opening a dedicated forum to allow all internet users, not just Facebook members, to read and post comments on Jodcast-related items, or on astronomy in general.

## 6. Advertising and Promotion

The Jodcast has not been promoted aggressively, nor have any advertising campaigns been conducted in order to increase the number of listeners. Advertising has been limited to the creation of a banner at the Jodrell Bank Observatory Visitor Centre, and a set of promotional postcards. The Jodcast has been advertised on several astronomy and science newsgroups, and by poster presentations made at the 2007 and 2008 UK National Astronomy Meetings. It appears that listeners to The Jodcast have became aware of the programme via web search engines, or word-of-mouth, or other direct referral. A Facebook ad campaign was recently started and has been driving around 10 – 15 people per day (limited by the maximum daily advertising budget) to the Jodcast website.

The Jodcast logo has been used on promotional items such as t-shirts and postcards. We have been selling promotional t-shirts via the Jodrell Bank Observatory Visitor Centre on an informal basis. Negotiations are ongoing with The University of Manchester to make on-line sales of these and similar items possible online.

### 7. External Involvement

We promote and welcome any other producer of astronomy-related audio who wishes to contribute their content to an episode of The Jodcast. Producing a podcast with regular and frequent episodes requires a significant time commitment. However, there are occasions where one-off events or opportunities arise for astronomy amateurs or professionals who may be interested in making an audio programme, but do not wish to enter into the commitment of producing a podcast. We operate The Jodcast as a channel for astronomy-related content, rather than a closed production.

We have also at times requested listeners to donate their time and skills toward improving The Jodcast. For example, Susan M. Lockwood, a music producer in Los Angeles, has composed introduction and incidental music for the video episodes of The Jodcast. All contributors to The Jodcast naturally get full credit on the corresponding episode page.

Related to the "open-source" nature of The Jodcast, all audio productions of The Jodcast are released under the Creative Commons Attribution-NonCommercial-ShareAlike 2.0 England and Wales Licence, which allows users to copy, distribute, display and perform

The Jodcast, and to make derivative works so long as The Jodcast is given credit, not used for commercial purposes, and any derivative works distributed under the same licence.

In 2007, a CDROM compilation was made of the best audio segments from the 2006 Jodcast season, and made available to any interested science teachers or education institution. Key words from the three main A-level curricula were linked to content on the CDROM, with the intention that interested science teachers might use Jodcast content to provide ideas for illustrating the real-life implementation of physical phenomena discussed in class. The intention was for teachers to use the CDROM content as inspiration for themselves, rather than as a teaching aid in itself. Feedback from teachers who used the CDROM showed that in most cases, teachers were more interested in products that could be immediately used as a teaching aid.

Again though its policy of open involvement, The Jodcast will seek more interaction with science teachers at local high schools. Scripting, recording and editing a piece of audio or video work for broadcast by The Jodcast is a potential project which could interest the science, English, IT and media departments of a high school.

## 8. Discussion

A key performance indicator is the number of listeners to The Jodcast, and we have seen the audience size increase from 300 listeners of the first episode to currently over 2800 listeners per episode. We seek to increase our audience size, but not through means which would compromise the production qualities and ethos of The Jodcast. We have received substantial feedback from regular listeners who praise the format and style of The Jodcast, as well as the level of required understanding. The Jodcast produces lengthly (around 60 minute) podcasts with minimal arbitrary stimulus (e.g. incidental music, sound effects). It is likely that these production qualities dissuade younger listeners. The Jodcast has identified a niche market, and to alter the show style now simply to increase audience numbers by sacrificing content would leave a disappointed audience. The Jodcast does aim to educate and inspire young people in particular, and for this reason, The Jodcast has extended into video podcasting.

### 9. The Future: Jodcast Video

The Jodcast was awarded a second STFC Small Award for the Public Engagement of Science. The majority of this award funding has been used to purchase a high-definition video camera (Sony HDR-FX1), a shotgun microphone and boompole (Rode). Adobe Premiere Pro CS3 was purchased to perform editing and post-production tasks.

Throughout 2008, filming has been ongoing, and the first episode released in September 2008, with further episodes becoming available at regular intervals.

The decision to include video products as part of The Jodcast was partly in an attempt to attract a younger audience. Feedback received as part of an informal survey conducted amongst physics teachers regarding The Jodcast audio programme typically noted that video, or other imagery is vital to capture and maintain the interest of a younger audience. Producing any video footage requires significantly more editing time compared to an audio product of the same duration. With this in mind, an upper limit on the duration of video Jodcast productions was set at 5 minutes. This consequently sets a limit on the depth to which any topic can be discussed.

Again, aware of the significant time commitment required for video production, The Jodcast hired two additional personnel for the video Jodcast project. Colin Stuart and Emily Fair both had substantial experience producing science video podcasts and expressed interest in working with The Jodcast.

To date, the footage for 9 video podcasts has been recorded, and the shows are in various stages of editing and post-production. Video shows are usually rendered at four resolutions: high and low Quicktime, Flash and full High Definition (1080i).

## 10. Summary

The Jodcast is a successful public outreach project, growing out of the enthusiasm and drive of a small number of dedicated science communicators. Starting without funding or expensive equipment, The Jodcast has increased its audience by presenting the science authentically: clearly, without sacrificing detail in order to achieve brevity, or to artificially stimulate or maintain interest through stylistic device.

The Jodcast is currently produced "in-house", with all writing, editing and recording done by team members. The quantity of Jodcast productions is currently sustainable, any increase is likely to require additional personnel.

Despite being a study renowned for dazzling images, astronomy can quite successfully be communicated through the medium of radio, or audio podcast. The scientists interviewed for The Jodcast have always been capable of explaining their research without reference to specific images or graphs. Over the course of The Jodcast project, professional educators have noted that video produc-tions are more likely to engage and enthuse a young audience. Rather than abandon the successful audio podcast, we decided to create video podcasts in addition to the audio Jodcast. The hour-long audio podcast, with its extended interviews and higher level of detail contrasts with the shorter, more easily assimilated video podcasts. By extending into video podcasting, we hope to engage with a younger audience.

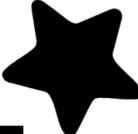